\documentclass{osa-article}
\usepackage{graphicx}
\usepackage{dcolumn}
\usepackage{bm}
\usepackage{braket}
\usepackage{xcolor}

\journal{osajournal}

\articletype{Research Article}

\begin{document}

\title{Efficient amplification of superpositions of coherent states using input states with different parities}

\author{Changhun Oh and Hyunseok Jeong}

\address{Center for Macroscopic Quantum Control, Department of Physics and Astronomy, Seoul National University, Seoul 08826, Korea}




\begin{abstract}
We investigate an experimentally feasible scheme for amplification of superpositions of coherent states (SCSs) in light fields. 
This scheme mixes two input SCSs at a 50:50 beam splitter and performs post-selection by a  homodyne detection on one output mode.
The key idea is to use two different types of SCSs with opposite parities for input states, which results in an amplified output SCS with a nearly perfect fidelity.
\end{abstract}

\section{Introduction}
Superpositions of coherent states (SCSs) in traveling light fields have been recognized as a good resource for quantum information processing such as
quantum teleportation \cite{teleport1,teleport2,teleport3}, quantum computation \cite{jeong2002, ralph2003}, quantum metrology \cite{ralph2002, munro2002, joo2011,joo2012} and fundamental tests of quantum mechanics \cite{schleich1991, wenger2003, jeong2003, stobinska2007, jeong2008, etesse2014}.
Their interesting features as macroscopic quantum superpositions have also been pointed out \cite{lee2011, jeong2015, frowis2015}.
Therefore, the importance of obtaining SCSs with sufficiently large amplitudes and high fidelities is twofold.
It is helpful for practical implementations of quantum information processing, and
it is also desirable  from the viewpoint of macroscopic quantum tests.
There have been numerous proposals \cite{lund2004, jeong2005, takeoka2007, marek2008, menzies2009, laghaout2013} and experimental implementations \cite{ourjoumtsev2006, nielsen2006, ourjoumtsev2007, takahashi2008, gerrits2010, nielsen2010, yukawa2013, miwa2014, huang2015, etesse2015, sychev2017} for the generation of SCSs.

There have been proposals for the amplification of small SCSs to large-size SCSs using measurement and post-selection.
A feasible  scheme using beam splitters, on-off photodetectors, and post-selection  was suggested  \cite{lund2004,jeong2005}.
A modified  setup using homodyne detection instead of the on-off detection for the amplification of SCSs was investigated \cite{takeoka2007, laghaout2013} and recently implemented in experiment \cite{sychev2017}.
In  Ref.~\cite{laghaout2013}, the authors proposed a setup for generation of a larger even SCS using two SCSs with the same parity as input states.
Using the proposed setup with two smaller odd SCSs, a larger even SCS was generated in an experiment~\cite{sychev2017}.
However, in comparison to the suggestion in Refs.~\cite{lund2004,jeong2005}, this approach using homodyne detection \cite{takeoka2007, laghaout2013} causes  the fidelity to degrade. In order to obtain sufficiently large SCSs, one may need to apply this amplification process for multiple times, and this is a formidable obstacle  particularly when initial SCSs are small.

In this paper, we investigate another set of input states which are composed of an odd SCS and an even SCS to overcome the obstacle.
First of all, we present and compare the performance of amplification scheme using two different pairs of input states by means of post-selection with homodyne detection. We then show that a nearly perfect fidelity can be attained using the pair of SCSs with opposite parities even when the amplitude of input SCSs is small whereas the fidelity that can be achieved by using a pair of small SCSs with the same parities is low. Thus, SCSs with opposite parities outperform those with same parities in the amplification procedure.

This paper is organized as follows.
In section \ref{sec2}, we present the setup for amplification and investigate the capability of the amplification procedure by calculating the fidelity of the output state heralded by homodyne outcome and probability to get the outcome. Then we present the probability to attain a target fidelity for different amplitude of input states.
We also compare two different sets of input states and show that a nearly perfect fidelity is always achievable with SCSs of opposite parities, which is not the case when SCSs with same parities are used.
Finally, we summarize our works in section \ref{sec3}.

\section{Amplification of Superposition of coherent states} \label{sec2}
In this section, we investigate the scheme that amplifies two SCSs into a larger SCS, where SCSs are defined by
\begin{equation}
\ket{\text{SCS}_\pm(\alpha)}\equiv\mathcal{N}_\pm(\alpha)(|\alpha\rangle\pm|-\alpha\rangle).
\end{equation}
Here, $\ket{\pm\alpha}$ are coherent states of amplitude $\pm\alpha$, and $\mathcal{N}_\pm(\alpha)=[2\pm2\exp(-2\alpha^2)]^{-1/2}$ are normalization factors. Note that $\ket{\text{SCS}_+} (\ket{\text{SCS}_-})$ is also called an even (odd) SCS because $\ket{\text{SCS}_+} (\ket{\text{SCS}_-})$ contains even (odd) number of photons only. We assume the amplitude $\alpha$ to be real without loss of generality throughout the paper, and drop the explicit argument $\alpha$ in the state representation for simplicity in following equations unless there is any confusion.

In the first part of the present section, we review the amplification scheme using two odd (even) SCSs and show the performance based on output fidelity and probability to obtain high fidelity outcomes.
We show that this scheme requires large amplitude SCSs for high fidelity output states.
In the second part, we investigate the amplification using two different types of SCSs with opposite parities in the same way.
Conclusively, we show that in this case nearly perfect amplification is possible regardless of the amplitude of input SCSs.

\begin{figure}[t]
                \centering
                \includegraphics[width=0.60\textwidth]{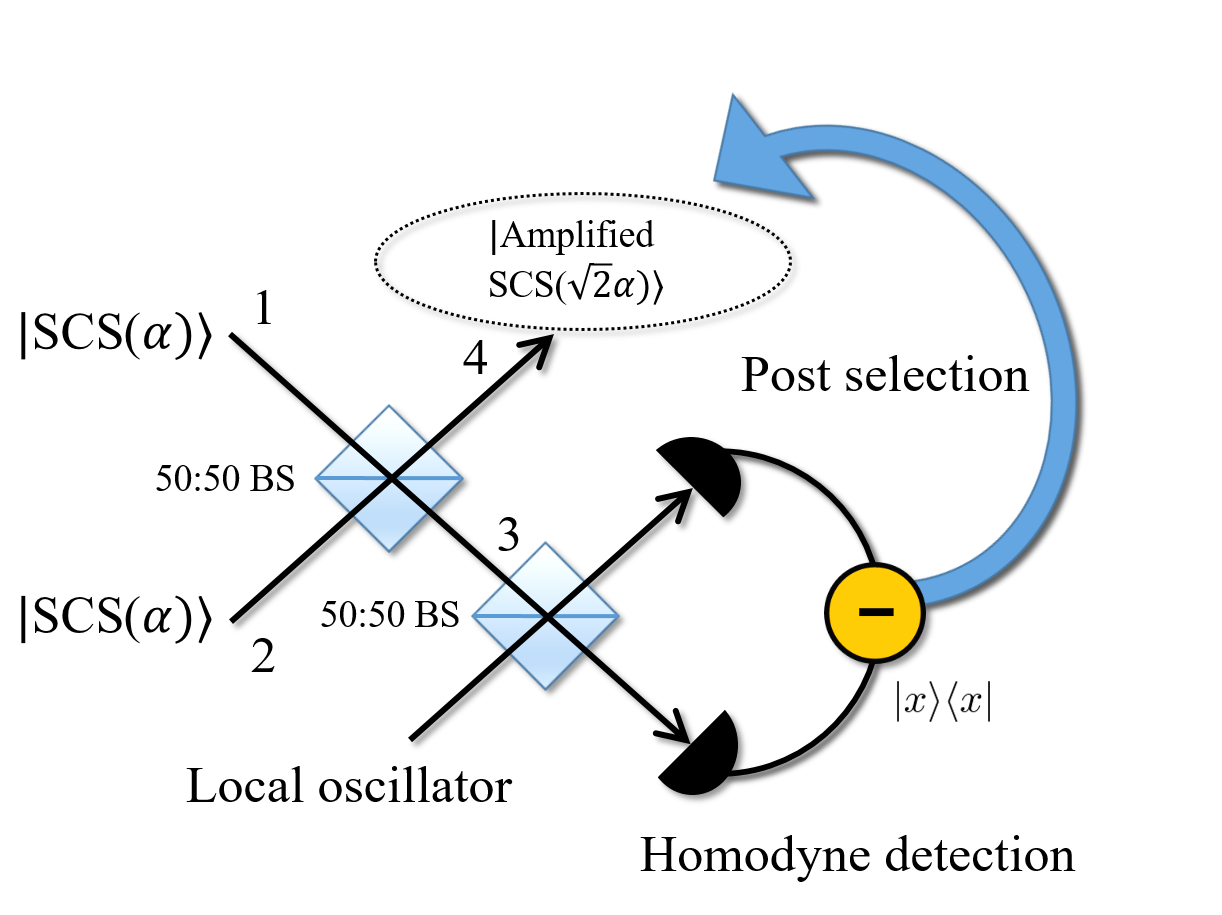}
                \caption{Setup for the amplification of SCSs. See the text for details.}
		\label{scheme}
        \end{figure}

\subsection{Amplifcation from two odd (even) SCSs to a larger even SCS}
\begin{figure}[t]
\includegraphics[scale=0.3]{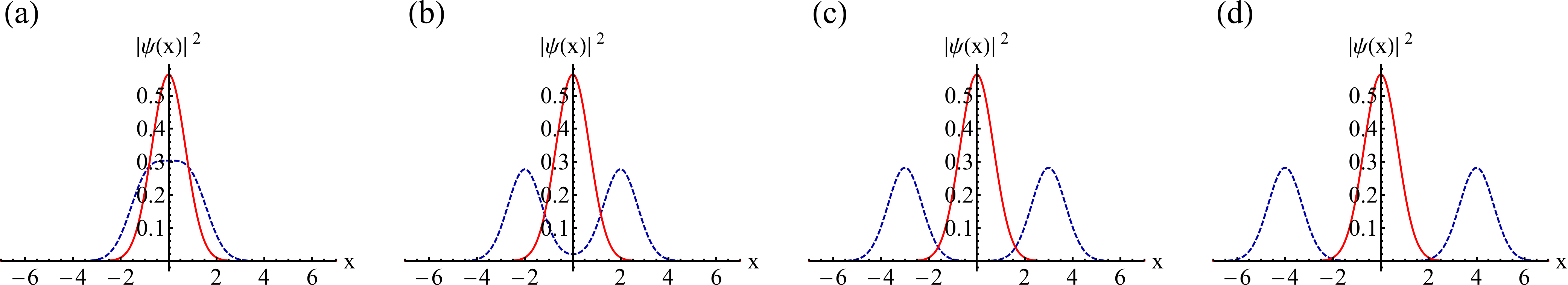}
\centering
\caption{Comparison of quadrature representation of the vacuum state (solid line) and even SCSs (dashed line) for different amplitude $\alpha$. (a) $\alpha=0.5$. (b) $\alpha=1.0$. (c) $\alpha=1.5$. (d) $\alpha=2.0$. In the cases of (a) and (b), $\psi_0(x)$ and $\psi_{\sqrt{2}\alpha}(x)+\psi_{-\sqrt{2}\alpha}(x)$ have large overlaps, which prevent amplification to high fidelity SCSs. On the other hand, the overlaps decrease as $\alpha$ increases so that high fidelity SCSs can be obtained (see (c) and (d)).}\label{homox1}
\end{figure}

We first present the amplification process with two odd SCSs \cite{laghaout2013, takeoka2007, sychev2017} (see Fig.~\ref{scheme}).
First of all, a pair of odd SCSs with same amplitude $\alpha$ are incident onto a 50:50 beam splitter as
\begin{align}\nonumber
|\text{SCS}_-\rangle_1|\text{SCS}_-\rangle_2 \xrightarrow{\text{BS}} \ket{0}_{3}(\ket{\sqrt{2}\alpha}_{4}+\ket{-\sqrt{2}\alpha}_{4})-(\ket{\sqrt{2}\alpha}_{3}+\ket{-\sqrt{2}\alpha}_{3})\ket{0}_{4},
\end{align}
where the normalization factor is omitted on the right hand side.
The $x$ quadrature is then measured by homodyne detection on the field of mode 3, where the $x$ quadrature operator is defined as $\hat{x}=(\hat{a}+\hat{a}^\dagger)/\sqrt{2}$ and $\hat{a}$ and $\hat{a}^\dagger$ are the annihilation and creation operators.
If the homodyne measurement outcome on mode 3 is $x=x_0$, the state on mode 4 is projected onto $\ket{\Psi(x=x_0)}_4$ with the probability density $p(x=x_0)$ such as
\begin{align}
\ket{\Psi(x=x_0)}_4&=\mathcal{N}_-^2\frac{\psi_0(x_0)(|\sqrt{2}\alpha\rangle_4+|-\sqrt{2}\alpha\rangle_4)-(\psi_{\sqrt{2}\alpha}(x_0)+\psi_{-\sqrt{2}\alpha}(x_0))|0\rangle_4}{\sqrt{p(x=x_0)}}, \label{wf0} \\
p(x=x_0)&=\mathcal{N}_{-}^4[(\psi_{\sqrt{2}\alpha}(x_0)+\psi_{-\sqrt{2}\alpha}(x_0))^2+\psi_0(x_0)^2\mathcal{N}_{+}(\sqrt{2}\alpha)^{-2} \nonumber \\ &\qquad\quad-4\psi_0(x_0)(\psi_{\sqrt{2}\alpha}(x_0)+\psi_{-\sqrt{2}\alpha}(x_0))e^{-\alpha^2}].
\end{align}
Here, $\psi_\alpha(x)\equiv\langle x|\alpha\rangle=\pi^{-1/4}\exp[-(x-\sqrt{2}\alpha)^2/2]$ is the quadrature representation of a coherent state. 
As seen in Eq. (\ref{wf0}), if the measurement outcome $x=x_0$ satisfies two conditions, (i) $\psi_{\sqrt{2}\alpha}(x_0)+\psi_{-\sqrt{2}\alpha}(x_0)= 0$ and (ii) $\psi_0(x_0)\neq 0$, the $\sqrt{2}$ times amplified even SCS would be obtained at mode 4.
In other words, if we post-select the output state according to the homodyne measurement outcome of the $x$ quadrature on the mode 3, we can obtain an amplified SCS on mode 4. 
In order to find the region where the above two conditions are satisfied, we compare $\psi_{\sqrt{2}\alpha}(x)+\psi_{-\sqrt{2}\alpha}(x)$ and $\psi_0(x)$ in Fig. \ref{homox1}.
The figure shows that for a large amplitude $\alpha$, the above conditions are approximately satisfied for $x_0$ around 0. Thus, in this case, we can expect the output state on mode 4 to be close to $\sqrt{2}$ times amplified even SCS.
On the other hand, when the amplitude $\alpha$ is too small such as $\alpha$=0.5, the above conditions cannot be satisfied for any $x$.
Thus, the preparation of large amplitude SCSs is necessary for the amplification for obtaining a high fidelity output state to the $\sqrt{2}$ times amplified even SCS.

\begin{figure}[t]
\hspace{0.3cm}
\includegraphics[scale=0.80]{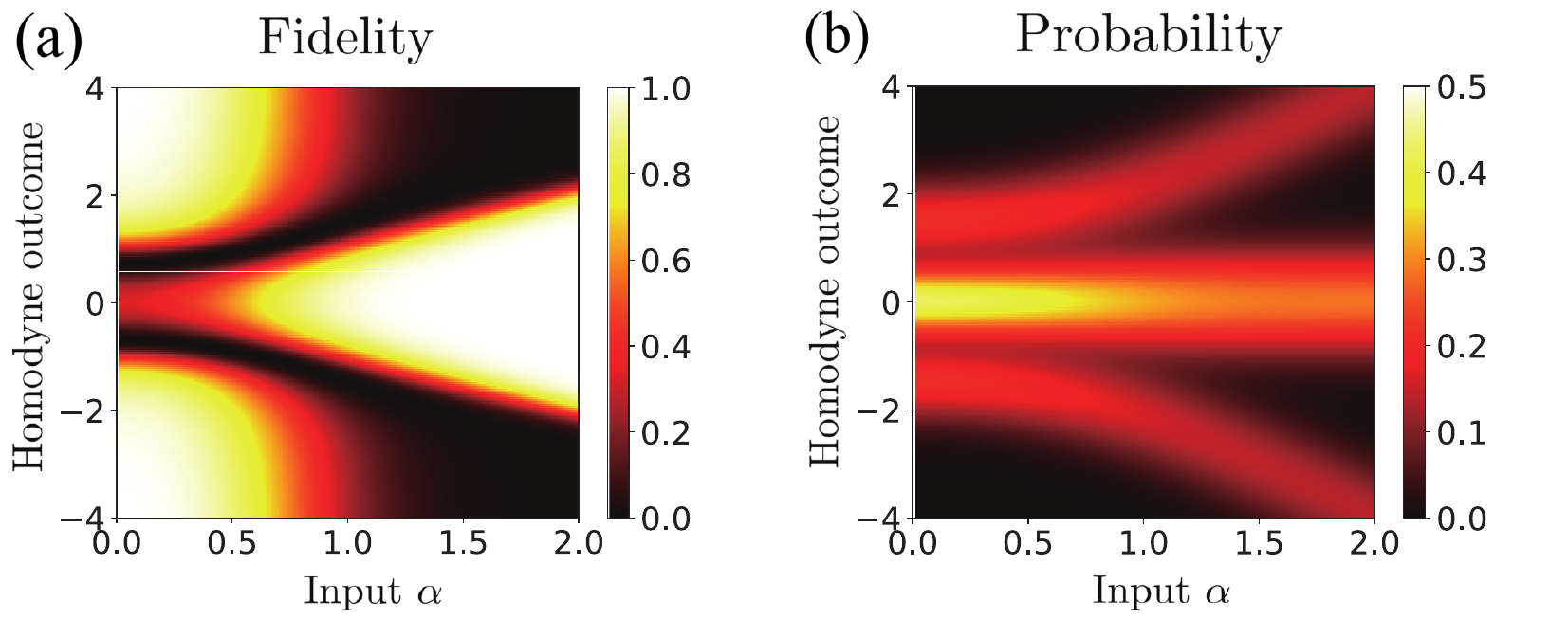}
\caption{(a) Fidelity between the outcome state and the ideal amplified SCS with different input amplitudes $\alpha$ and homodyne outcomes $x_0$. (b) Probability density of getting homodyne outcomes $x_0$ for different $\alpha$. While there are three regions where high fidelity output states can be obtained, the probablity to get left upper and lower regions is very low, which is shown in (b). Since the fidelity is not large enough when the homodyne outcome is around 0 and input $\alpha$ is low, the achievable fidelity is low when input states with low amplitudes $\alpha$ are used.}\label{den1}
\end{figure}

In order to analyze more quantitatively, we plot the fidelity to $\sqrt{2}$ times amplified even SCS when the input SCS has the amplitude $\alpha$ and the measurement outcome is $x=x_0$, and the probability density to obtain the outcome, which are shown in Fig. \ref{den1}. Here, the fidelity is defined as 
\begin{align}
F_+(\alpha,x_0)&=|\langle \text{SCS}_+(\sqrt{2}\alpha)|\Psi(x=x_0)\rangle|^2 \nonumber \\
&=\frac{\mathcal{N}_{-}^4}{p(x=x_0)}\bigg[\frac{\psi_0(x_0)}{\mathcal{N}_+(\sqrt{2}\alpha)}-2 e^{-\alpha^2}\mathcal{N}_+(\sqrt{2}\alpha)(\psi_{\sqrt{2}\alpha}(x_0)+\psi_{-\sqrt{2}\alpha}(x_0))\bigg]^2. 
\end{align}
Fig. \ref{den1} (a) shows that there are two regions where the high fidelity output state can be obtained : (i) large input amplitude $\alpha$ with measurement outcome $x_0$ around 0 and (ii) the small input amplitude $\alpha$ with large measurement outcome $|x_0|$. However, since the probability to get the region (ii) is very small, which is shown in Fig. \ref{den1} (b), we focus only on the region (i) from now on. The region (i) implies that if one prepares large SCSs as input states, such as $\alpha\geq 1.0$, and obtains the homodyne outcome around 0, the resultant fidelity to amplified SCS will be close to 1.

\begin{figure}[b]
\includegraphics[scale=0.68]{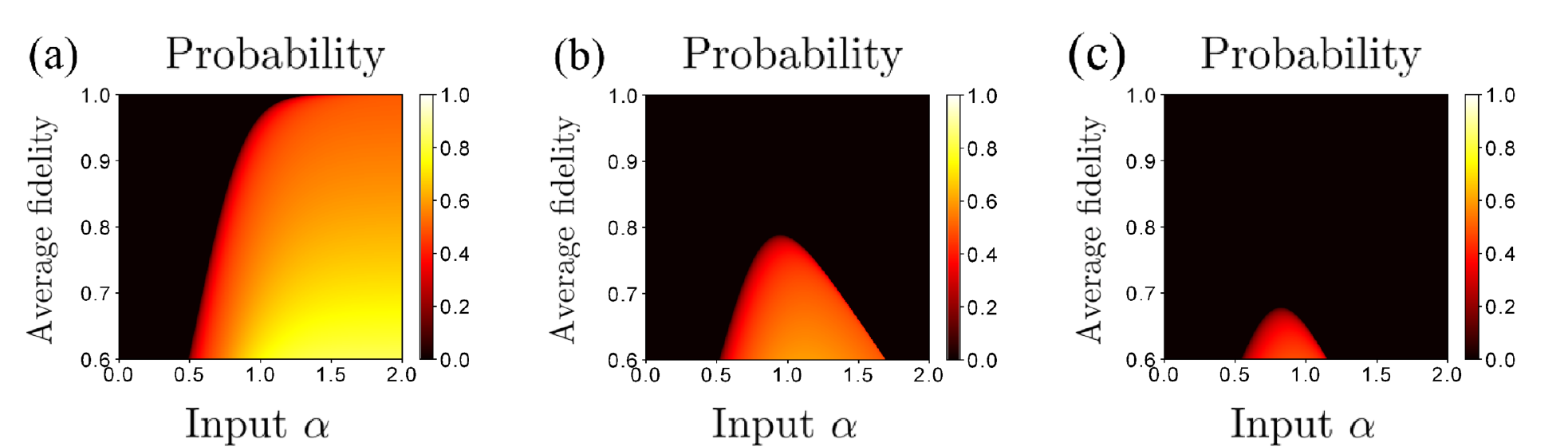}
\caption{Probability to get the target fidelity with input SCS with amplitude $\alpha$ (a) without photon-loss, (b) with 10$\%$ photon-loss, and (c) with 20$\%$ photon-loss on input states. When the amplitude of input SCSs is low, the probability to get high fidelity SCSs is extremely low, which can be overcome by using different types of SCSs as we will see in the following subsection. When photon-loss occurs, it is difficult to amplify large amplitude input SCSs to a high fidelity SCS because they are more fragile to loss than small amplitude SCSs.}\label{sucprob1}
\end{figure}

Now, we assume that one sets a window $[-x_0, x_0]$ of post-selection so that the output state is selected if the homodyne outcome is in the window.
Then, on average, the resultant fidelity on average will be 
\begin{align}
\mathcal{F}_+(\alpha, x_0)=\frac{\int_{-x_0}^{x_0} p(x)F_+(\alpha,x)dx}{\int_{-x_0}^{x_0} p(x)dx}.
\end{align}
The probability of getting a target fidelity on average when the input amplitude is $\alpha$ is shown in Fig. \ref{sucprob1}. Note that we have used average fidelity on a region $[-x_0, x_0]$.
The figure shows that the probability of getting high fidelity outcome with small amplitude input states is extremely small.
Thus, the amplification to high fidelity output state is possible only for high amplitude input states.
Therefore, nearly perfect amplification with a pair of odd SCSs requires the preparation of large amplitude of input SCSs.

It is important to consider experimental imperfections. Here, we assume photon loss on input states. If we consider loss of $r^2$ by a beam splitter model, input SCSs will evolve as
 \cite{walls1994}
\begin{align}
|\text{SCS}_\pm\rangle\langle \text{SCS}_\pm|\rightarrow |t\alpha\rangle\langle t\alpha|+|-t\alpha\rangle\langle -t\alpha|\pm e^{-2r^2\alpha^2}(|t\alpha\rangle\langle -t\alpha|+|-t\alpha\rangle\langle t\alpha|),
\end{align}
where $t^2+r^2=1$ and $r^2$ is a photon-loss rate and the normalization constant is omitted.
Let us denote the state after the 50:50 beam splitter in the main setup with input SCSs under photon loss  as $\rho(\alpha)$ and the state after homodyne outcome $x_0$ as $\tilde{\rho}(\alpha,x_0)={}_3\langle x_0|\rho(\alpha)|x_0\rangle_3$.
The probability density of obtaining  homodyne outcome $x_0$ and the fidelity of the output state with homodyne outcome $x_0$ to an amplified ideal SCS are written as
\begin{align}
p(x=x_0)&=\text{Tr}[\tilde{\rho}(\alpha, x_0)] \\
F_+(\alpha,x_0)&=\langle \text{SCS}_+(\sqrt{2}\alpha)|\tilde{\rho}(\alpha, x_0)|\text{SCS}_+(\sqrt{2}\alpha)\rangle.
\end{align}
We then use the fidelity of the output state to an ideally amplified even SCS to numerically calculate the average fidelity with a window $[-x_0, x_0]$.
In the following subsection, we will show that degradation of performance of the scheme due to photon loss can be overcome to some extent with different types of input SCSs.

So far we have presented the amplification scheme that uses two odd SCSs to obtain a larger even SCS. It can be easily checked that two even SCSs input case yields a similar result.
In the following subsection, we show that different input states result in a nearly perfect amplification without large amplitude SCSs.

\begin{figure}[b]
\includegraphics[scale=0.3]{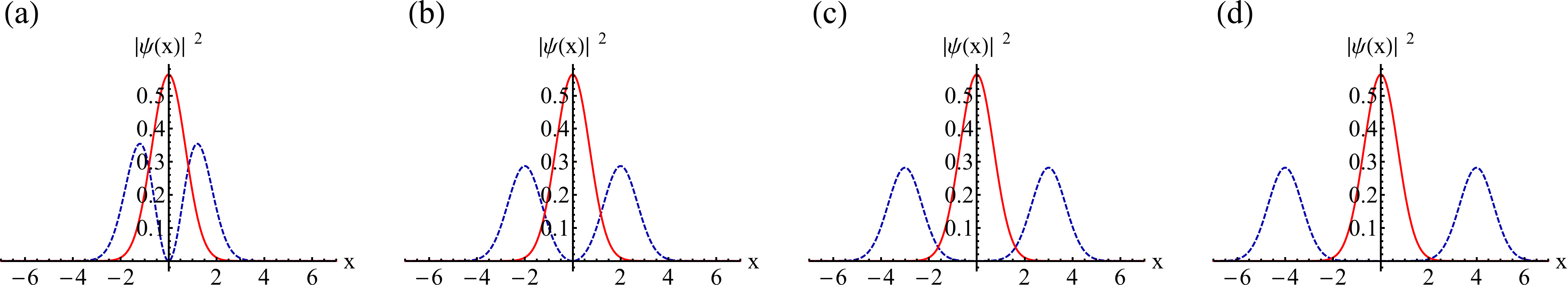}
\centering
\caption{Comparison of quadrature representation of the vacuum state (solid line) and odd SCSs (dashed line) for different amplitude $\alpha$. (a) $\alpha=0.5$. (b) $\alpha=1.0$. (c) $\alpha=1.5$. (d) $\alpha=2.0$. Figures (c) and (d) are very similar to Fig. 2 (c) and (d). High fidelity amplified SCS can be obtained for large $\alpha$. Remarkably, even when $\alpha$ is small, nearly perfect amplification is possible because $\psi_0(x)$ and $\psi_{\sqrt{2}\alpha}(x)+\psi_{-\sqrt{2}\alpha}(x)$ does not overlap around $x=0$ regardless of $\alpha$.}\label{homox2}
\end{figure}

\subsection{Amplification from an even SCS and an odd SCS to a larger odd SCS}
In this subsection, we change the input states to an odd SCS and an even SCS and analyze the result in the same procedure.
Similarly, an odd SCS and an even SCS are incident onto a 50:50 beam splitter as
\begin{align}\nonumber
|\text{SCS}_+\rangle_1|\text{SCS}_-\rangle_2 \xrightarrow{\text{BS}}\ket{0}_{3}(\ket{\sqrt{2}\alpha}_{4}-\ket{-\sqrt{2}\alpha}_{4})+(\ket{\sqrt{2}\alpha}_{3}-\ket{-\sqrt{2}\alpha}_{3})\ket{0}_{4},
\end{align}
where the normalization factor is omitted. Again, we measure the $x$ quadrature of the beam on the mode 3 using homodyne detection. After measuring $x=x_0$ on mode 3, the state on mode 4 is projected onto
\begin{align}
\ket{\Psi(x=x_0)}_4=\frac{\mathcal{N}_+\mathcal{N}_-}{\sqrt{p(x=x_0)}}[\psi_0(x_0)(|\sqrt{2}\alpha\rangle_4-|-\sqrt{2}\alpha\rangle_4)+(\psi_{\sqrt{2}\alpha}(x_0)-\psi_{-\sqrt{2}\alpha}(x_0))|0\rangle_4],\label{proj2}
\end{align}
with the probability density,
\begin{align}
p(x=x_0)=\mathcal{N}_+^2\mathcal{N}_-^2[\psi_0(x_0)^2 \mathcal{N}_-(\sqrt{2}\alpha)^2+(\psi_{\sqrt{2}\alpha}(x_0)-\psi_{-\sqrt{2}\alpha}(x_0))^2].
\end{align}
As seen from Eq. (\ref{proj2}), if we measure $x=x_0$ satisfying two conditions, (i) $\psi_{\sqrt{2}\alpha}(x_0)-\psi_{-\sqrt{2}\alpha}(x_0)= 0$ and (ii) $\psi_0(x_0)\neq 0$, an amplified odd SCS with an amplitude $\sqrt{2}\alpha$ would be obtained at mode 4.
In order to find the region where the above two conditions are satisfied, we again compare $\psi_{\sqrt{2}\alpha}(x)-\psi_{-\sqrt{2}\alpha}(x)$ and $\psi_0(x)$, which is shown in Fig. \ref{homox2}. The figure shows that the results are similar with those of the previous case for large amplitude $\alpha$. However, in this case, even for small $\alpha$ such as $\alpha=0.5$, we can still find some region where a larger odd SCS is obtained because $\psi_{\sqrt{2}\alpha}(0)-\psi_{-\sqrt{2}\alpha}(0)$ is always zero. In other words, in contrast to the two odd SCSs input case, the amplification of SCS with an odd SCS and an even SCS is always possible for arbitrary $\alpha$.

\begin{figure}[t]
\hspace{0.3cm}
\includegraphics[scale=0.80]{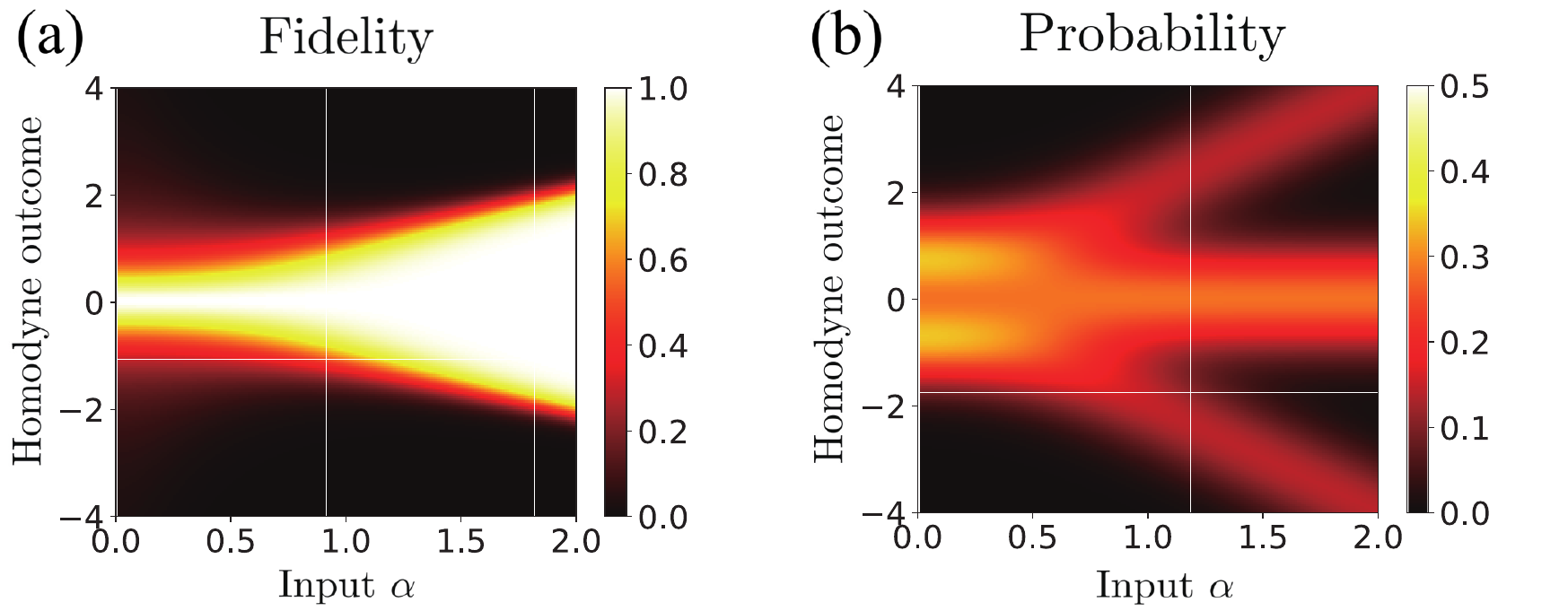}
\caption{(a) Fidelity between the outcome state and the ideal amplified SCS with different input amplitudes $\alpha$ and homodyne outcomes $x_0$ for input states with different parities. (b) Probability density of getting homodyne outcomes $x_0$ for different $\alpha$. A nearly perfect amplification is possible regardless of input $\alpha$. Especially, small amplitude SCSs can be used for nearly perfect amplification, which is different from Fig. 3.}\label{den2}
\end{figure}

\begin{figure}[b]
\includegraphics[scale=0.70]{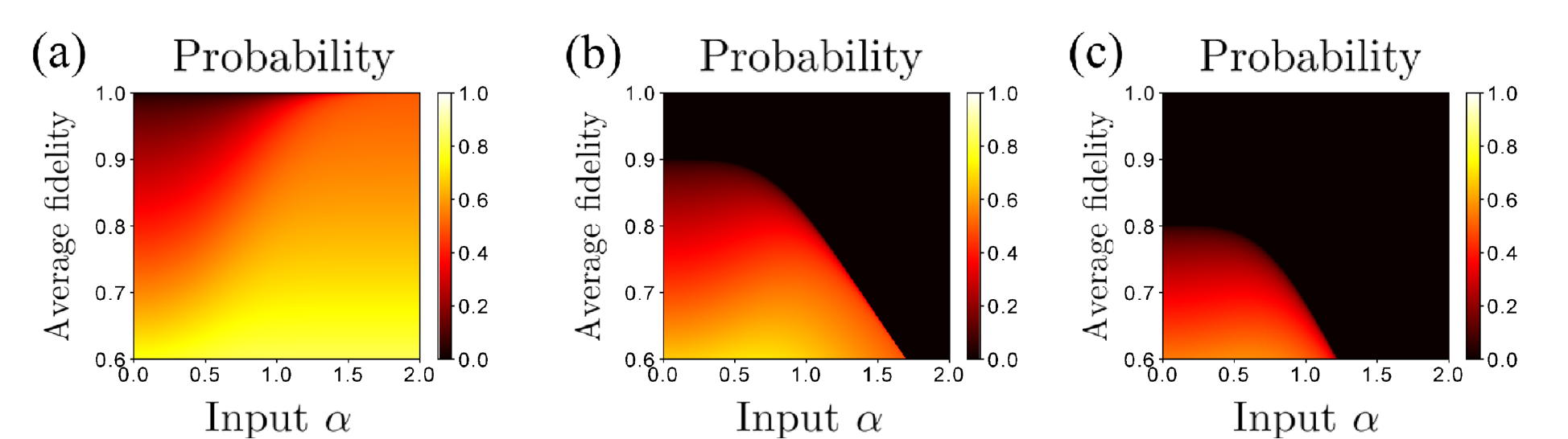}
\caption{Probability to get the target fidelity with input SCS with amplitude $\alpha$ (a) without photon-loss, (b) with 10$\%$ photon-loss, and (c) with 20$\%$ photon-loss on input states. The advantage over the previous case can be seen for small $\alpha$. In this case, high fidelity SCSs can be generated even when input $\alpha$ is small. Again, it is more difficult to amplify large amplitude input SCSs to a high fidelity SCS under photon loss.}\label{sucprob2}
\end{figure}

We analyze the fidelity of the output state to the $\sqrt{2}$ times amplified odd SCS. When we measure $x=x_0$ on mode 3, the fidelity of the output state on mode 4 to a larger odd SCS is given as
\begin{align}
F_{-}(\alpha,x_0)=|\langle \text{SCS}_-(\sqrt{2}\alpha)|\Psi(x=x_0)\rangle|^2=\frac{\mathcal{N}_+^2\mathcal{N}_-^2}{\mathcal{N}_-(\sqrt{2}\alpha)^2}\frac{\psi_0(x_0)^2}{p(x=x_0)}. 
\end{align}
The fidelity for different $\alpha$ and given measurement outcomes and the probability of getting the outcome are shown in Fig. \ref{den2}.
As the figure shows, conditioning the homodyne measurement outcome around $0$ results in approximately perfect amplification for arbitrary amplitude of input states.
As input amplitude $\alpha$ increases, a window $[-x_0, x_0]$ where we get high fidelity output state gets broader.

Again, we plot the probability of getting a target fidelity on average with setting the window $[-x_0, x_0]$, which is shown in Fig. \ref{sucprob2}.
Here, the average fidelity on a region $[-x_0,x_0]$ is written as
\begin{align}
\mathcal{F}_-(\alpha, x_0)=\frac{\int_{-x_0}^{x_0} p(x)F_-(\alpha,x)dx}{\int_{-x_0}^{x_0} p(x)dx}.
\end{align}
First of all, when the amplitude of input states is large, such as $\alpha\geq 1.5$, the figure is very similar to the previous case which is shown in Fig. \ref{sucprob1}.
In other words, the amplification from large SCSs input to $\sqrt{2}$ times amplified SCS with high fidelity is possible for both cases.
However, an important difference from the case with a pair of SCSs with same parities is that the probability of getting high fidelity with small amplitude of input states is much larger, while the probability of attaining high fidelity output state with the input pair of two small odd SCSs is negligible.
Moreover, the advantage becomes more important in a realistic situation where photon-loss occurs on input SCSs, which is shown in Figs. \ref{sucprob2}(b) and (c). Compared to the previous setup, as shown in Figs. \ref{sucprob1}(b) and (c), even when photon-loss is considered, our scheme has larger probabilities to attain high fidelity SCSs especially for low amplitude input SCSs.
Thus, the pair of SCSs with different parities is advantageous in that arbitrary amplitude SCSs can be used for the amplification.

\begin{figure}[t]
\centering
\includegraphics[scale=0.6]{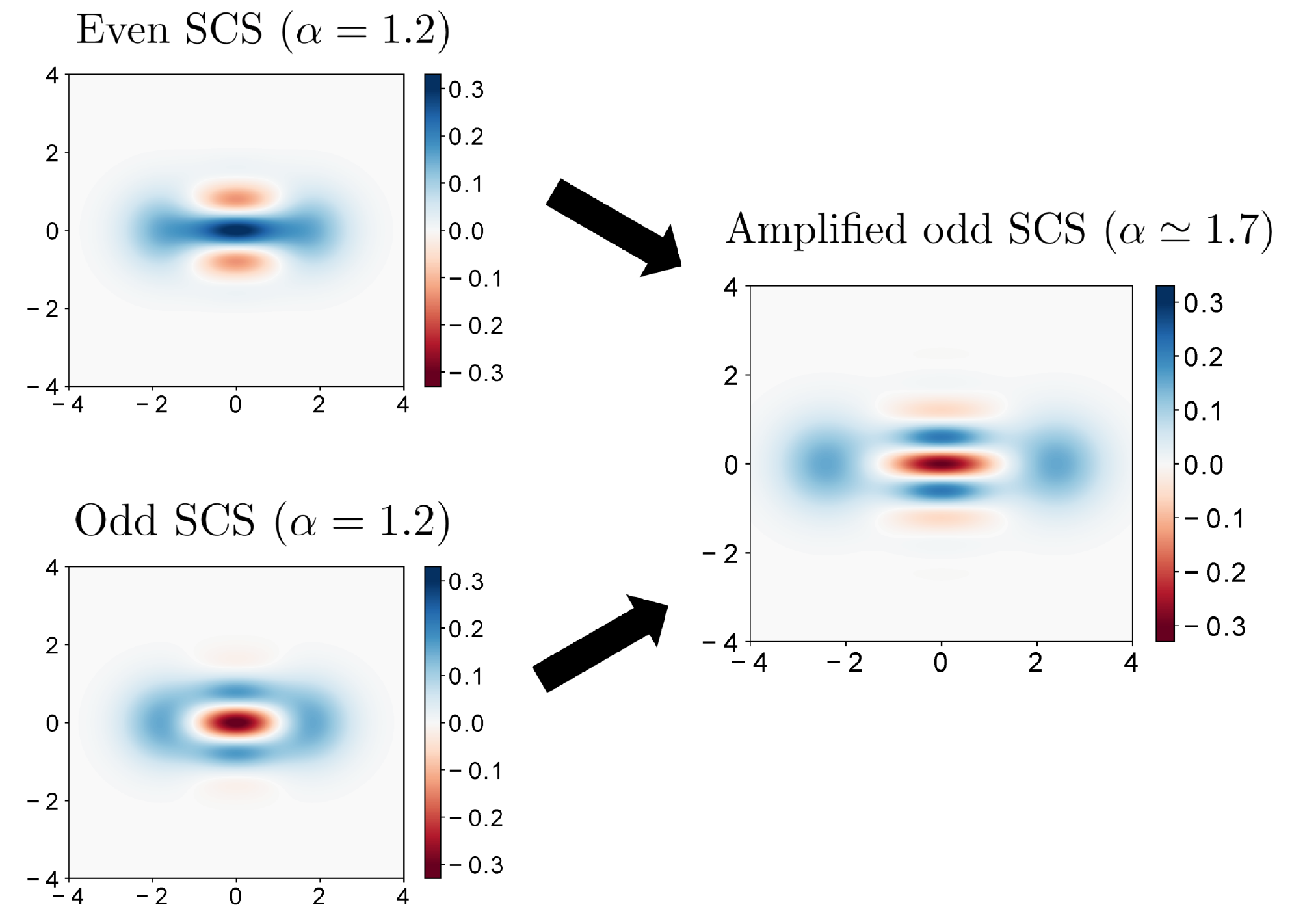}
\caption{Wigner functions of input SCSs of amplitude $\alpha=1.2$ (left) and of amplified SCS of $\alpha\simeq1.7$ (right). Comparing the input odd SCS and the resultant odd SCS, one can observe the change of the interference fringe of odd SCS. We set the windows of homodyne outcome as $[-1, 1]$, which gives the success probability of $43\%$ and fidelity of 0.9754.}\label{wigner}
\end{figure}

Figure~\ref{wigner} shows an example of a SCS amplified out of an odd SCS and an even SCS. 
The figure clearly shows the increase of both the amplitude and the frequency of interference fringes that is an evidence of the increase of the nonclassical property \cite{lee2011}.

\section{Conclusion}\label{sec3}
We have investigated a conditional amplification scheme for free-traveling optical SCSs. 
This scheme simply mixes two input SCSs at a 50:50 beam splitter and performs post-selection by means of homodyne detection on one of the two output modes.
The output state heralded by an appropriate homodyne outcome is close to the $\sqrt{2}$ amplified SCS.
We considered two different pairs of input states for the amplification.
One is two odd SCSs, and the other is an odd SCS and an even SCS. 
We have shown that both pairs of input states with large amplitude enable the amplification to high fidelity output state.
However, more importantly, we found that a nearly perfect fidelity can be obtained using two different types of SCSs with opposite parities for input states, whereas large amplitude SCSs are necessary for a pair of SCSs with the same parities.
The advantage of using two different SCSs over the same SCSs is crucial for generating sufficiently large amplitude SCSs from small amplitude SCSs because multiple times of imperfect amplification will degrade the fidelity of the final output state to the large SCS.
Since two different SCSs input enables a nearly perfect amplification for arbitrary amplitude of input SCSs, it can be used iteratively to generate a nearly perfect SCS with large amplitude.

Experimental implementation of our scheme for amplification of SCSs requires preparation of input SCSs.
A well-known technique to generate an approximate odd SCS is to subtract a single photon from a squeezed vacuum state \cite{lund2004, jeong2005, ourjoumtsev2006, nielsen2006, marek2008, takahashi2008, gerrits2010}.
An approximate even SCS can be generated using a number state and a homodyne detection \cite{ourjoumtsev2007}.
Altenatively, once we prepare an odd SCS, one can generate an approximate even SCS by subtracting a single photon from the odd SCS, $\hat{a}(|\alpha\rangle\pm|-\alpha\rangle)\propto(|\alpha\rangle\mp|-\alpha\rangle)$.
It is well known that the photon subtraction can be realized using a high-transmissivity beam splitter and a photon detector \cite{ourjoumtsev2006}.
This means that an approximate even SCS can be produced by subtracting two photons from a squeezed vacuum state \cite{marek2008}, which has been realized in experiments \cite{takahashi2008, gerrits2010}.
In fact, more generally, subtracting arbitrary number of photons results in a well-approximated SCS with some degrees of additional squeezing \cite{marek2008}.
There is another scheme  to mix two squeezed vacuum states at an asymmetric beam splitter and perform photon-number resolving detection on one of two output modes, which results in an even or odd SCS depending on the measurement outcome \cite{huang2015}.
In this context, it is worth noting that there is a study for optimization over input Gaussian states with photon-number resolving detection to generate approximate non-Gaussian states including SCSs \cite{menzies2009}. On the other hand, interconversion between single photons and SCSs can be also employed, which has been realized in experiment \cite{miwa2014}. These ideas may be combined with our scheme for better performance.
In experiments, these approximate SCSs may be used as seeds of the amplification scheme to obtain larger SCSs.
Of course, the approximate SCSs used as input SCSs will cause degradation of fidelities for output SCSs obtained by the amplification process. It will be worthwhile to analyze our amplification scheme in realistic setups in a more detailed and rigorous way  as a future work.

\section{Acknowledgements}
This work was supported by the National Research Foundation of Korea (NRF) through a Grant funded by the Korean government (MSIP) (Grant No. 2010-0018295), and the Korea Institute of Science and Technology Institutional Program (Project No. 2E27800-18-P043).






\end{document}